# Impact of Solar Activity on the Ionosphere/Thermosphere during Geomagnetic Quiet Time for CTIPe and TIE-GCM

A. Fitzmaurice, M. Kuznetsova, J.S. Shim, V. Uritsky

## Abstract

This study examines the impact that solar activity has on model results during geomagnetic quiet time for the ionosphere/thermosphere models: the Coupled Thermosphere Ionosphere Plasmasphere Electrodynamics Model (CTIPe) and the Thermosphere-Ionosphere-Electrodynamics General Circulation Model (TIE-GCM). Using varying F10.7 flux values as a measurement of solar activity, the models were run over a two-day period with the constant parameters Kp= 2, n= 3 $cm^3$, and v= 400 km/s. Four F10.7 values (70, 110, 150, and 190) were selected based off of the average F10.7 values for geomagnetic quiet days across the current solar cycle. Our analysis of the model results showed that solar activity has the greatest effect on TEC output and the least effect on hmF2 and O/N2 output. Overall, TIE-GCM output tends to be higher than CTIPe, however, inconsistent values across the two-day CTIPe v3.1 output suggests that this model needs a longer warm-up period. When visually compared against observed data, CTIPe v3.2 performs the best, although all models greatly overestimated values for TEC. Analyses of models during geomagnetic quiet time are important in establishing a baseline which can be compared against storm data. An example of this was completed at the end of the study, using storm data obtained from Millstone Observatory on March 17, 2013. In this analysis, we found that TIE-GCM using the Weimer model to calculate the high-latitude electric potential provided more accurate results. Continued research in this area will be useful for understanding the impact of geomagnetic storms on the thermosphere/ionosphere.

## 1 Introduction

The thermosphere/ionosphere system is a highly variable, complex layer of the Earth's upper atmosphere. The thermosphere, named for its high temperatures, lies within altitudes between ~100 and 600 km above the surface of the Earth and is comprised of O, $O_2$, and $N_2$. Overlapping the thermosphere, along with portions of the mesosphere and exosphere, is the ionosphere (~60-1000 km), composed of partially ionized gas formed by solar radiation. Discovered in 1901 by Marconi, the ionosphere plays a significant role in radio signal transmission due to its ability to refract shortwave signals, allowing transmissions to be bounced off of the ionosphere in order to travel further distances. The rate at which a signal is refracted is highly dependent on the electron density, which can fluctuate due to changing solar activity and disturbances in the Earth's magnetic field. This study analyzes the effects that varying levels of solar activity have on electron density, and other related ionospheric parameters, during the 2013 spring equinox on geomagnetic quiet days at Millstone Hill Observatory (42.6 N, 71.5 W) and how this is presented by some common coupled ionosphere/thermosphere models. Section 2 reviews past related studies on this topic in addition to providing information on the models used. Section 3 describes the study of average F10.7 values for geomagnetic quiet days across the current solar cycle. Model results are discussed and compared against an empirical model and observed data in Section 4. Quiet data and storm data are compared in Section 5 and conclusions as well as topics of future research are presented in Section 6.

## 2 Background

Because of the large impact on radio transmissions and GPS signals, many studies have been conducted in recent years to research the effects of solar activity on the ionosphere. While long term trends are now predominately shaped by the rising levels of CO2 in the atmosphere [*Lastovicka et al.*, 2012], patterns in the ionosphere have been shown to follow recurring solar events such as varying EUV flux levels from active regions appearing and disappearing with the sun's 27 day rotation period [*Kutiev et al.*, 2012]. This is due to the strong correlation between electron density and solar radiation, so having a measurable index of solar activity is extremely important in studying and forecasting the ionosphere. A popular index such as this is the F10.7 solar radio flux. Taken at Dominion Radio Astrophysical Observatory in Penticton, British Columbia, Canada, the F10.7 flux is a measurement of the total radio emissions at the wavelength 10.7 cm from the entire solar disk for a one hour period. It is measured in "solar flux units (sfu)" which are equal to $10^{-22}$ W m$^{-2}$ Hz$^{-1}$. Three measurements are made every day at 1700, 2000, and 2100 UT. As discussed in *Tapping* [2013], there can be undersampling errors because of short changes in flux caused by solar activity such as flares, however, *Tapping and Charrois* [1994] found that flux measurements made at a single time of day were within two solar flux units of the daily average 95% of the time. It should be noted that while ionospheric and thermospheric parameters show a linear relationship with EUV radiation, there is a non-linear relationship between these parameters and the F10.7 flux, particularly for high flux values.

Though solar activity is an incredibly important factor in shaping the ionosphere, it is only one of many drivers which control ionospheric parameters. Because of this, a variety of coupled ionosphere/thermosphere models have been developed in recent years, many of which can be found at NASA's Community Coordinated Modeling Center (CCMC). The primary focus of this study are two physics-based, first principles models: the CTIPe [*Millward et al.*, 2001] and the TIE-GCM [*Qian et al.*, 2013]. TIE-GCM has two versions; one takes the interplanetary magnetic field and solar wind parameters to calculate the high-latitude electric potential using the *Weimer* [2005] model, while the other uses Kp index and the By component of the interplanetary magnetic field for the *Heelis et al.* [1982] model. Two versions of CTIPe are used in this study: version 3.1 and version 3.2.

## 3 Average F10.7 Values Across the Current Solar Cycle

To begin the study, F10.7 values had to be selected to be used as input for the models. This was done by recording the averages for all quiet days within a two month interval (Feb. 15-Apr. 15 for spring, May 15-July 15 for summer, Aug. 15-Oct. 15 for fall, and Nov. 15-Jan. 15 for winter) beginning in 2008 until present day. Thresholds for Kp, Dst, and AE were used to define quiet days (see Table 1). Data for Dst and AE was retrieved from Kyoto's World Data Center for Geomagnetism while data for Kp and F10.7 was retrieved from NOAA's National Geophysical Data Center.

Flux values were divided into three bins: low ($\leq$120), medium (120<f$\leq$180), and high ($\geq$180). The averages for each bin are presented in Figure 2. Our study has shown that even on geomagnetically quiet days, solar activity can be high. To demonstrate this, four values of F10.7 were chosen for the models: 70, 110, 150, and 190.

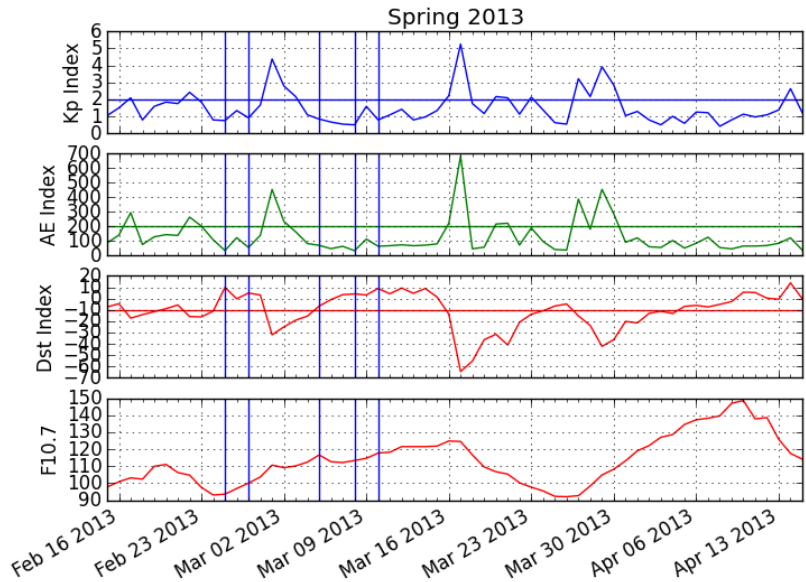

**Figure 1.** Kp, AE, Dst, and F10.7 for the 2013 spring equinox. Horizontal lines represent each index's threshold. Vertical lines represent the five quietest days. Plots like this were made for every season from 2008 to 2016.

**Table 1.** Quiet day thresholds for Kp, AE, and Dst.

| Index | Threshold |
|-------|-----------|
| Kp | $\geq 2$ |
| AE | $\geq 200$ nT |
| Dst | $\leq$ -10 nT |

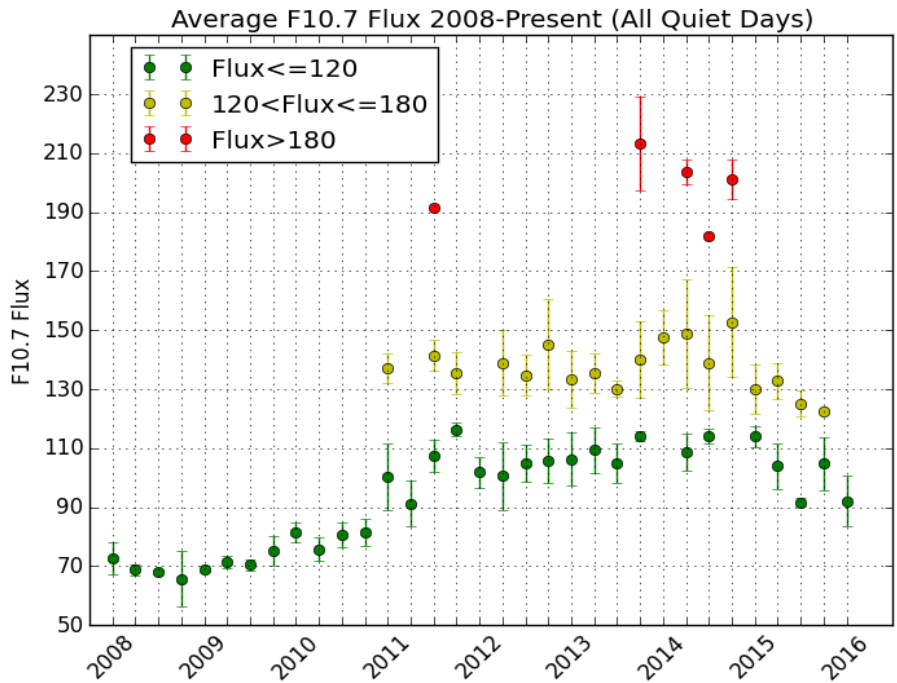

**Figure 2.** Average F10.7 values for each season from 2008 to spring 2016. Error bars represent the standard deviation of the data. As the solar cycle reaches its maximum, flux values as well as variation increase. Points where flux reached the high level are summer 2011, winter 2013, summer 2014, fall 2014, and winter 2014.

## 4 CTIPe and TIEGCM Analysis

### 4.1 Solar wind parameters for Kp = 2

Because several of the models used solar wind parameters instead of Kp as input, the solar wind conditions for Kp = 2 were found using the coupled solar wind/magnetosphere equations developed by *Newell et al.* [2008]:

$$Kp = 0.05 + 2.244 \times 10^{-4} \frac{d\Phi_{MP}}{dt} + 2.844 \times 10^{-6} n^{\frac{1}{2}} v^2 \tag{1}$$

$$\frac{d\Phi_{MP}}{dt} = v^{\frac{4}{3}} B_T^{\frac{2}{3}} \sin^{\frac{8}{3}}(\frac{\theta_C}{2}) \tag{2}$$

Here, *n* is density, *v* is the velocity, $B_T$ is the interplanetary magnetic field, and $\theta_C$ is the clock angle. For our study, we used $n = 3$ cm$^{-3}$ and $v = 400$ km/s. The corresponding $B_T$ and clock angle are shown in Figure 3. The parameters chosen for the study are highlighted in the provided table.

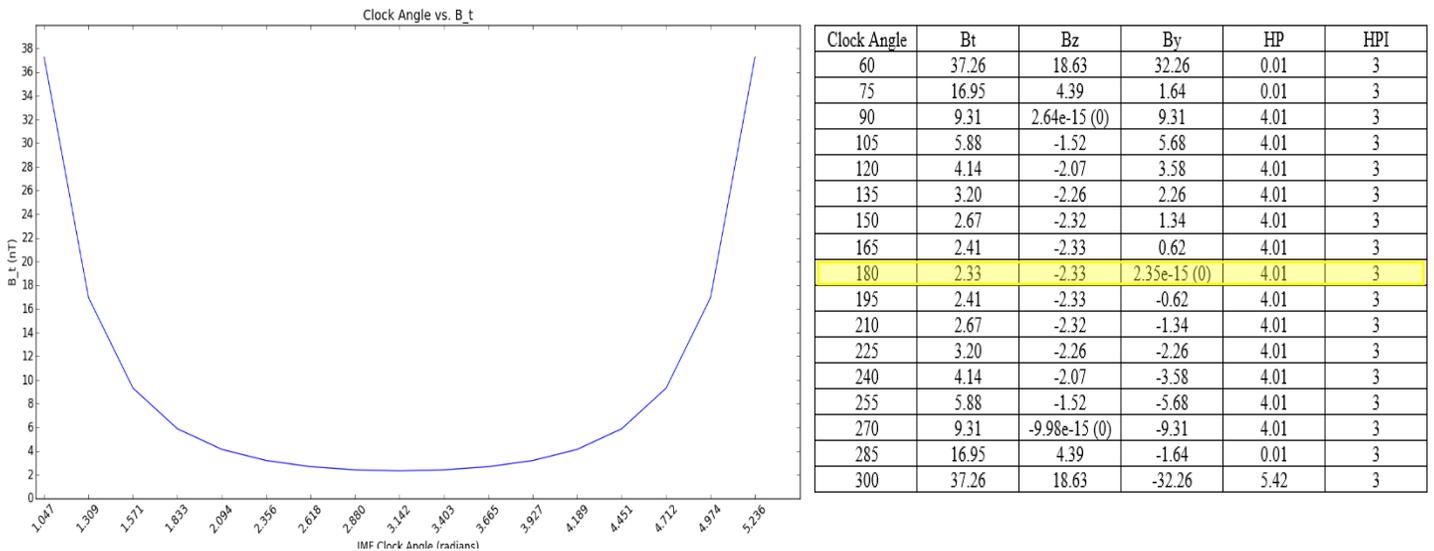

| Clock Angle | Bt | Bz | By | HP | HPI |
|---|---|---|---|---|---|
| 60 | 37.26 | 18.63 | 32.26 | 0.01 | 3 |
| 75 | 16.95 | 4.39 | 1.64 | 0.01 | 3 |
| 90 | 9.31 | 2.64e-15 (0) | 9.31 | 4.01 | 3 |
| 105 | 5.88 | -1.52 | 5.68 | 4.01 | 3 |
| 120 | 4.14 | -2.07 | 3.58 | 4.01 | 3 |
| 135 | 3.20 | -2.26 | 2.26 | 4.01 | 3 |
| 150 | 2.67 | -2.32 | 1.34 | 4.01 | 3 |
| 165 | 2.41 | -2.33 | 0.62 | 4.01 | 3 |
| 180 | 2.33 | -2.33 | 2.35e-15 (0) | 4.01 | 3 |
| 195 | 2.41 | -2.33 | -0.62 | 4.01 | 3 |
| 210 | 2.67 | -2.32 | -1.34 | 4.01 | 3 |
| 225 | 3.20 | -2.26 | -2.26 | 4.01 | 3 |
| 240 | 4.14 | -2.07 | -3.58 | 4.01 | 3 |
| 255 | 5.88 | -1.52 | -5.68 | 4.01 | 3 |
| 270 | 9.31 | -9.98e-15 (0) | -9.31 | 4.01 | 3 |
| 285 | 16.95 | 4.39 | -1.64 | 0.01 | 3 |
| 300 | 37.26 | 18.63 | -32.26 | 5.42 | 3 |

**Figure 3.** Clock angle (x-axis) versus interplanetary magnetic field (y-axis) for Kp = 2, $n = 3$ cm$^{-3}$, and $v = 400$ km/s. The parameters chosen for the study are $\theta_C = 180°$, Bt = 2.33 (By = Bx = 0, Bz = -2.33), Hemispheric Power (HP) = 4.0, and Hemispheric Power Index (HPI) = 3.

### 4.2 Comparing model results against each other

Before comparing the model results against the observed data, they were analyzed on their own to identify the differences between outputs for different flux values and different models. Every model was run with the same parameters for a total of two days with data every 15 minutes. While CTIPe v3.2, TIEGCM+Weimer, and TIEGCM+Heelis (Figure 4) all show excellent consistency between values for the first and second days, the results from CTIPe v3.1 (Figure 5) are different for values at the same time with the same parameters between the first and second days. This difference is most likely due to an insufficient amount of warm-up period for the CTIPe v3.1 model. Because of this error, when model results are compared to the observed data, the results from CTIPe v3.1 are not used.

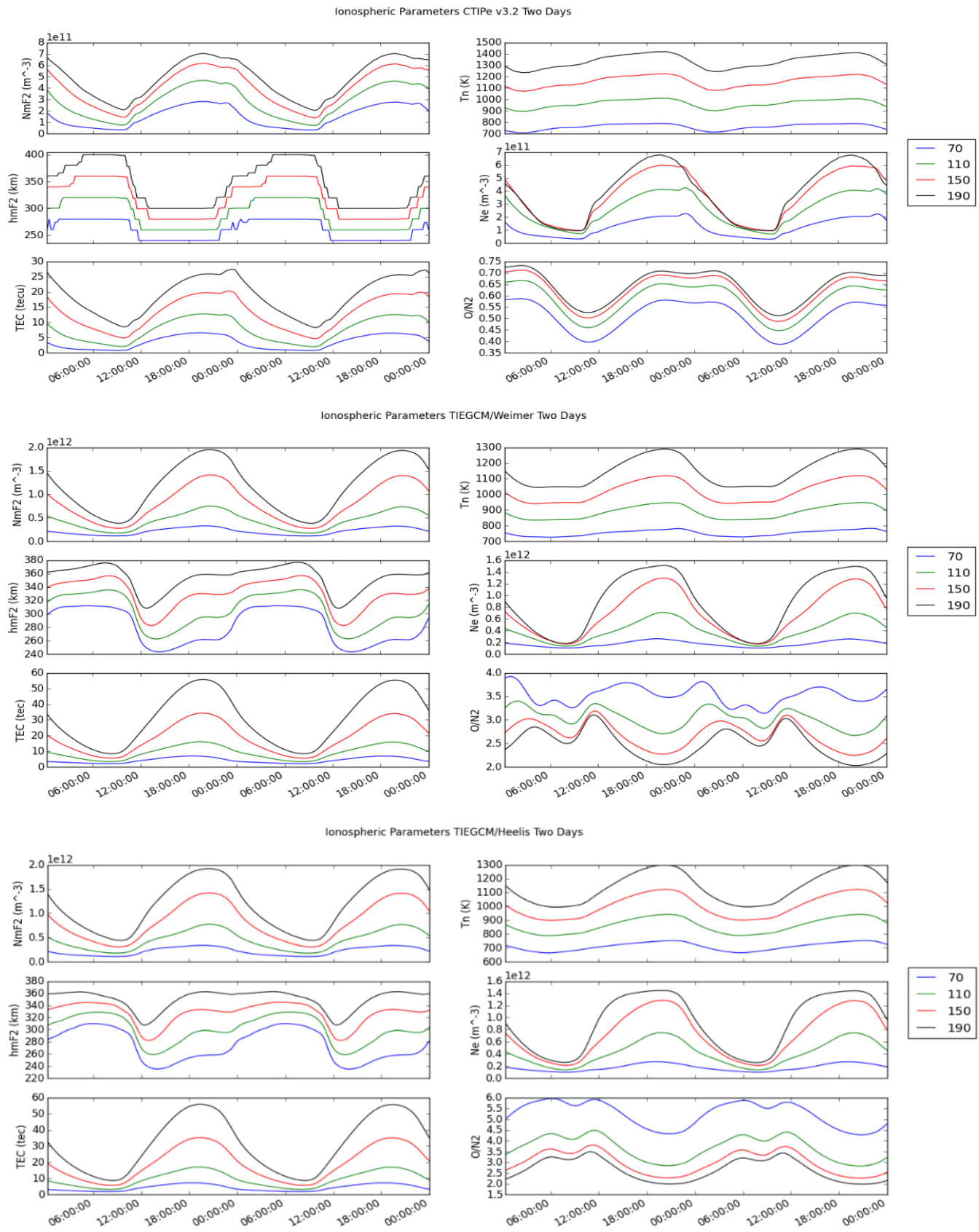

**Figure 4.** Full model results over the two day period for CTIPe v3.2 and both TIEGCM models. Parameters are (beginning in top left moving clockwise) NmF2, Tn at 300 km, Ne at 300 km, O/N2 at 300 km, TEC, and hmF2.

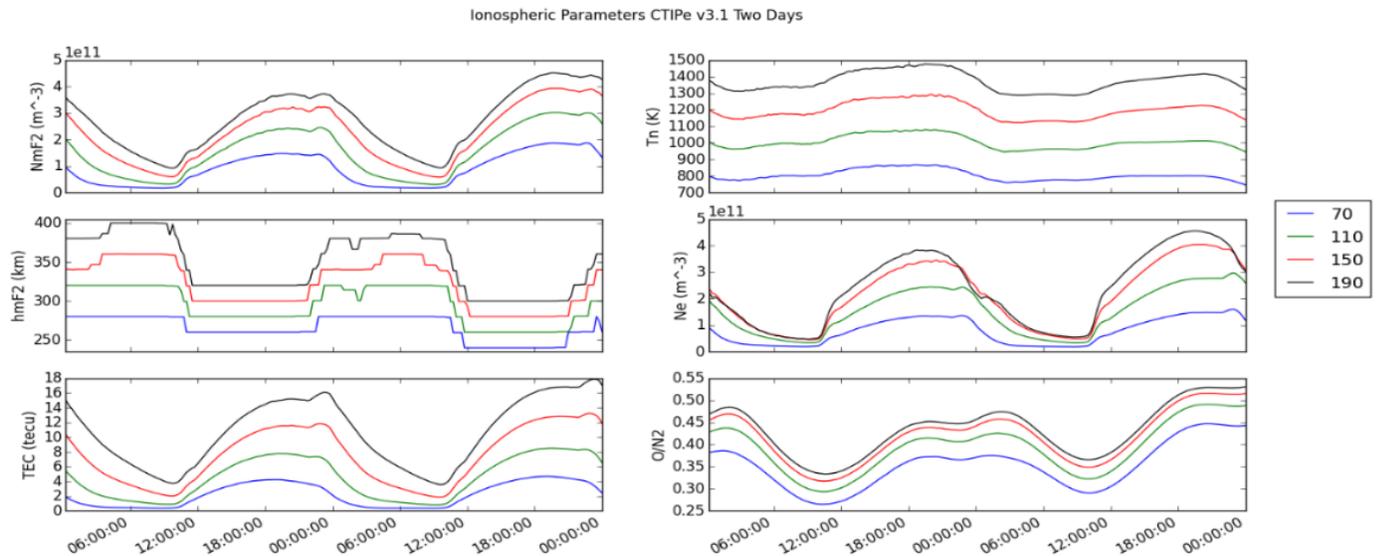

**Figure 5.** Full model results for CTIPe v3.1. Due to an improper warm-up period the output values change across the two days, despite the input parameters remaining constant.

Percent differences between data from different flux values were found to determine dependence on F10.7 and linearity. For the two versions of CTIPe (Figure 6), percent differences were highest for TEC and lowest for O/N2. Increased differences in a majority of the parameters occur between 0:00 and 12:00 UT (local nighttime/morning). Percent differences from Tn appear to be the most constant throughout the day. While NmF2, hmF2, TEC, and Tn all appear to be linear with F10.7, Ne and O/N2 show clear nonlinearity, especially for high flux values.

For the two versions of TIEGCM (Figure 7), percent differences were once again highest for TEC but lowest for hmF2. While most of the scales remain similar to those of the CTIPe models, the maximum percent difference from TEC decreased by ~1000% and the maximum from Ne increased by several hundred percent. Another major difference between the two models is that percent differences increase in many of the parameters between 12:00 and 0:00 UT (local daytime/early evening) instead of during local nighttime like the CTIPe models. Trends in linearity remain the same as the CTIPe models.

### 4.3 Models compared to empirical model and observed data

Because of the scarcity of available data, in addition to comparing the models to observed data, they were also compared to an empirical model developed by the MIT Haystack Observatory, as these models tend to be more accurate than physics-based models. Data from the empirical model is only available for NmF2. Comparisons to each flux value are presented in Figure 8.

A curve of observed data from 6 UT to 23:45 UT was obtained by compiling together data from 2009-03-24, 2009-03-25, and 2009-04-08. Data from 6 UT to 11:45 UT is from 03-25. Data from 12:15 UT to 13:15 UT and 20:15 UT to 23:45 UT is from 03-24. Data from 13:30 UT to 20:00 UT is the average of data from both 03-24 and 04-08. Data is available for NmF2, hmF2, TEC, and Ne at 300 km and was retrieved from the Madrigal online database. On every day used for observed data, Kp = 2 and F10.7 = 70. Comparisons for just F10.7 = 70 are available in Figure 9. Comparisons for every flux are available in Figure 10.

Because of the quality of the observed data, RMS analysis was not performed and instead the models are compared to the data qualitatively. For both the empirical model and the observed data, TIEGCM appears to have the best fit overall for NmF2, however, CTIPe v3.2 is closer to the observed data during the local afternoon. For the remainder of the parameters, CTIPe v3.2 provides the best fit to the data, however, it should be noted that all models slightly overestimate hmF2 during the local daytime. Results for TEC are furthest from the observed data for all models and are greatly overestimated during the majority of the day. For all models, the results for F10.7 = 70 fit best to the observed data.

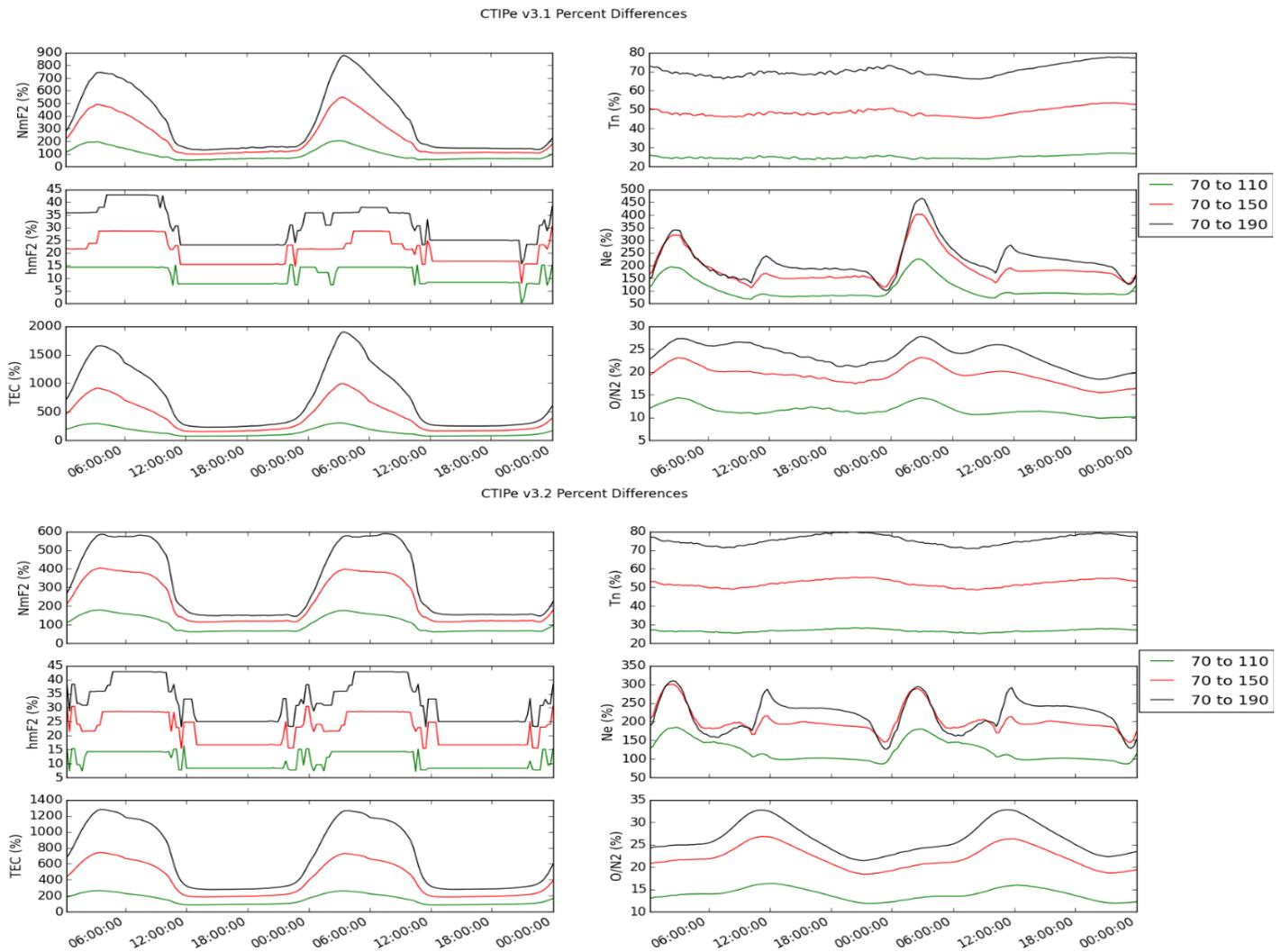

**Figure 6.** Percent differences between data from different flux values for the CTIPe models. Differences were found between each flux and the minimum (70).

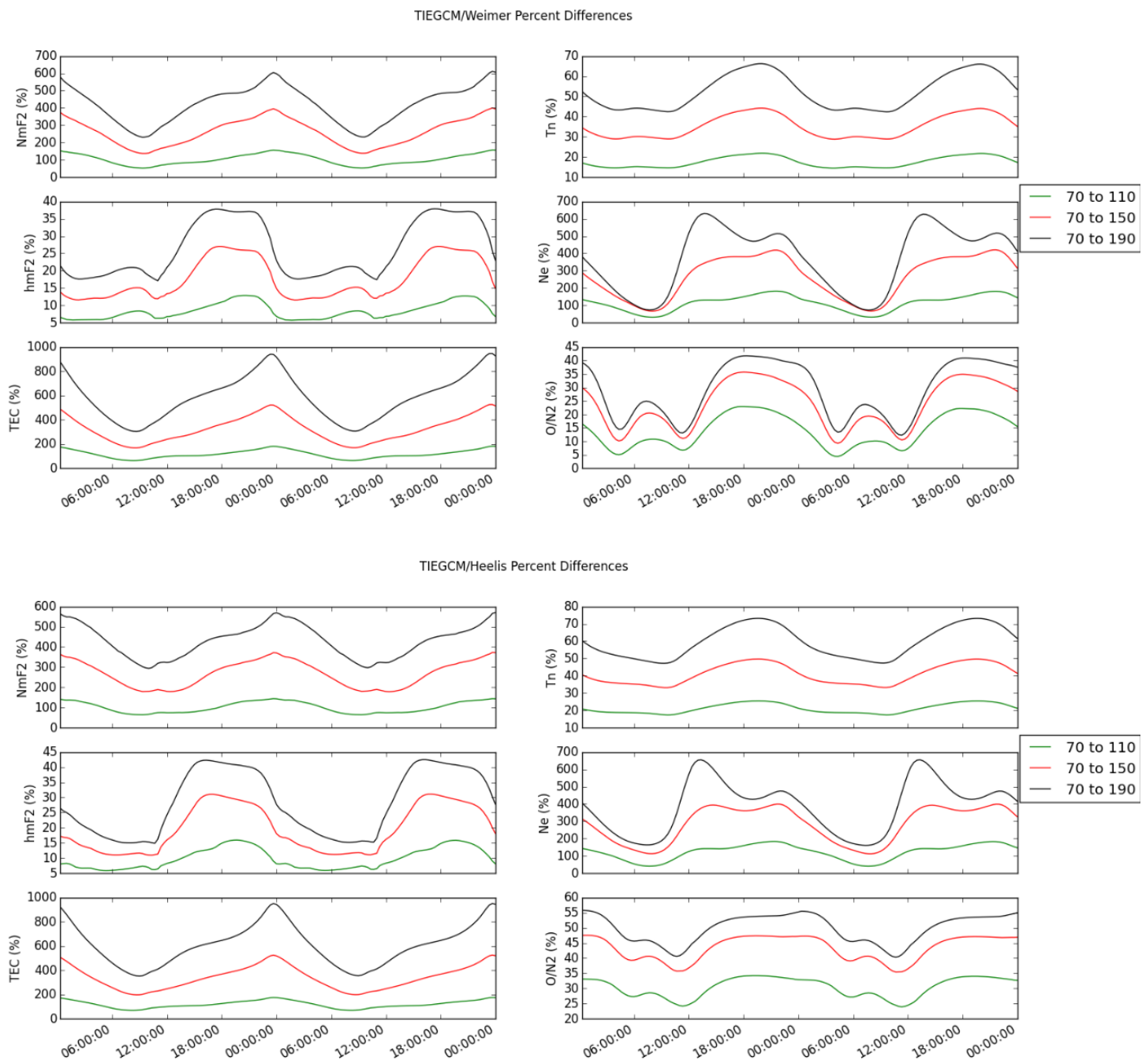

**Figure 7.** Percent differences between data from different flux values for the TIEGCM models. Differences were found between each flux and the minimum (70).

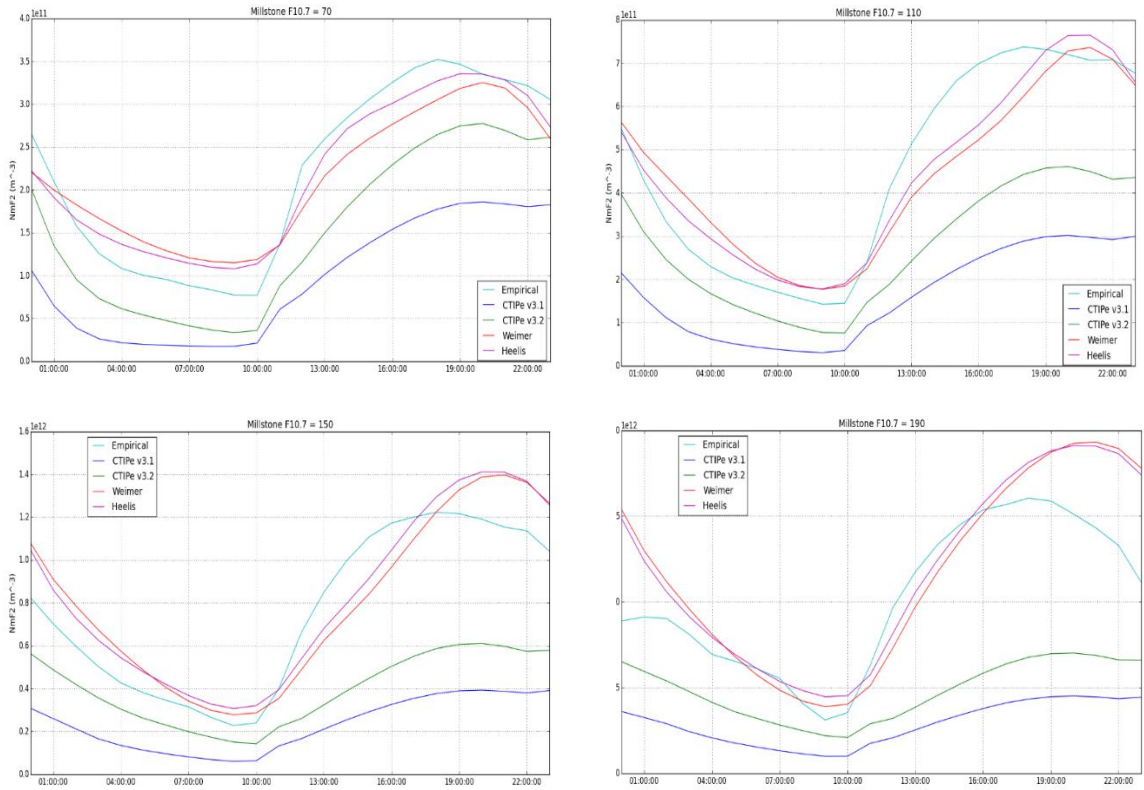

**Figure 8.** Comparisons between all models and the empirical model for every flux value from 0:00 to 23:00 UT with data every 1 hour. For every flux value, the TIEGCM models fit better to the empirical model than the CTIPe models, however, for high flux values, the TIEGCM models become much hotter at night than the empirical model.

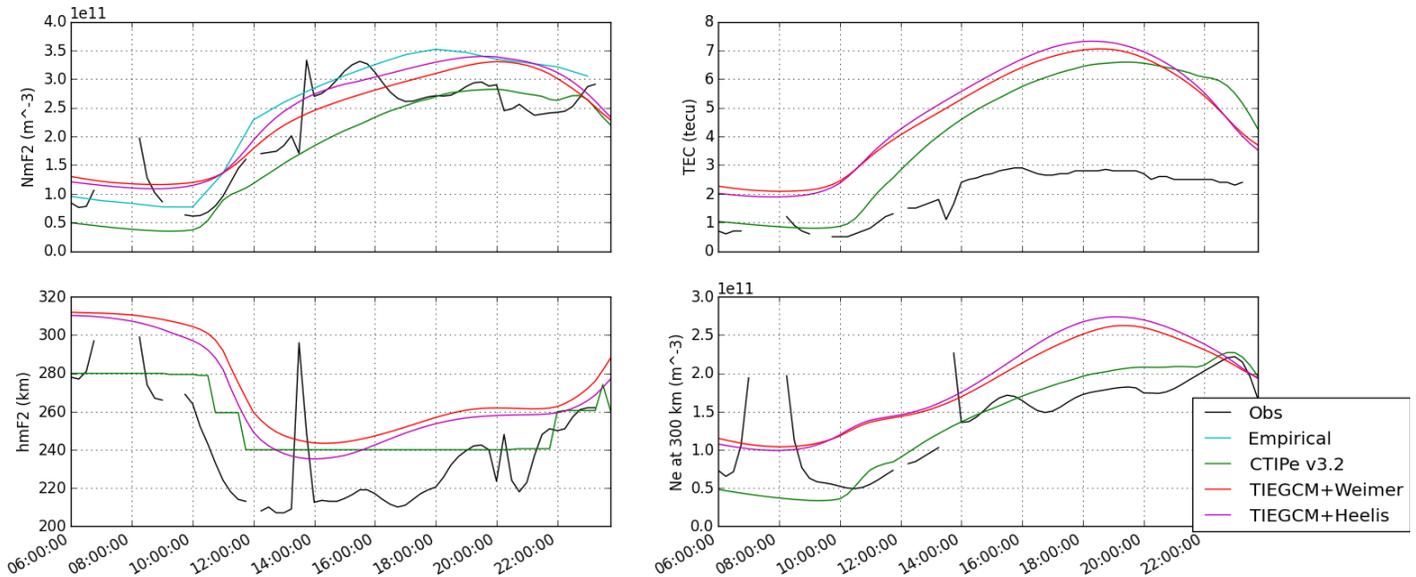

**Figure 9.** Models compared to observed data at Millstone Hill. CTIPe v3.2 (green) has the best fit to the observed data for all parameters except for NmF2, where TIEGCM+Weimer (red) shows the best agreement. All of the models greatly overestimate the data for TEC (top right) for all parts of the day except just before local dawn where CTIPe v3.2 appears to be quite close to the observed data.

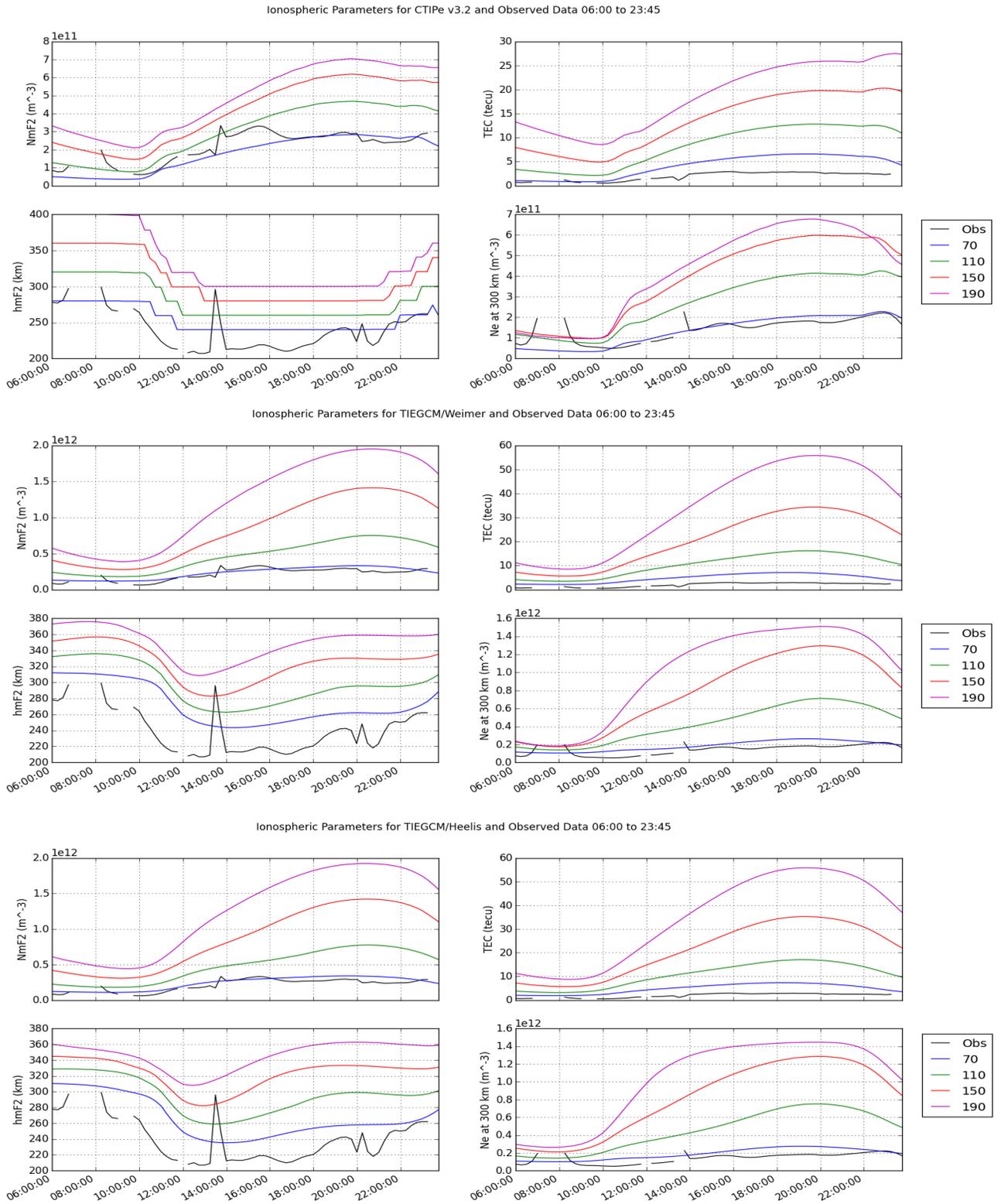

**Figure 10.** Comparisons of each model to the observed data for every flux value. For all models and all parameters, F10.7 = 70 results fit best to the observed data, however, many of the parameters are overestimated, especially in the TIEGCM models.

## 5 Quiet Data Compared to Storm Data

To complete the study, geomagnetic quiet data was compared to geomagnetic storm data, both observed and produced by the models. For this portion of the study, only two models were used to produce storm data: TIEGCM using Weimer and CTIPe v3.2. The storm analyzed here occurred on March 17, 2013.

For CTIPe, storm data results were run for five days, 03-15 to 03-19, with data every 15 minutes. For TIEGCM, storm data results were run for four days, 03-16 to 03-19, also with data every 15 minutes. Plots of storm results compared to quiet time results are shown in Figure 11. While the two models show similarities for hmF2 and Tn at 300 km, there are large differences between results for NmF2, TEC, Ne at 300 km, and O/N2 at 300 km. Whereas TIEGCM shows increases in these values due to the storm, CTIPe results decrease compared to the quiet time data.

Once again, observed data was retrieved from the Madrigal online database. Data was available for NmF2, hmF2, TEC, and Ne at 300 km from 03-16 at 16:15 UT to 03-17 at 22:30 UT, however, there is a large data gap at the beginning of the storm from 06:15 UT to 14:15 UT on 03-17. Observed data on its own is presented in Figure 12. There were two methods used to find the difference between quiet data and the observed storm data. The first, used the average of all the models (excluding CTIPe v3.1) at F10.7 = 110 (real F10.7 was ~126). The second, used the empirical quiet data at F10.7 = 110 as the baseline.

Model data was compared to observed data across four days: 03-16 to 03-19 (Figure 13 and Figure 14). In addition to comparing the data on its own, differences between quiet data and storm data are also compared. Because of the missing data, it is difficult to fully understand how the values compare, however, it does appear as though the available data includes the peak of the storm's ionospheric effects.

For the parameters NmF2 and Ne at 300 km, it is clear that TIEGCM+Weimer is the better model, though it does overestimate the effects for both. hmF2 results for both TIEGCM and CTIPe are comparable and show a good correlation with the observed data. Once again, TIEGCM greatly overestimates the effect on TEC, while CTIPe shows a much better fit.

The results for the comparison of the differences are similar. TIEGCM is obviously the better fit for NmF2 and Ne at 300. For NmF2, the model difference is slightly higher than the difference using the model average, but slightly lower than the difference using the empirical data. For Ne, the model results slightly overestimate the difference at the beginning of the storm, but are nearly perfect during the peak. hmF2 differences are not as comparable between the two models as the data itself and it appears as though TIEGCM fits slightly better than CTIPe. For TEC, CTIPe has the best fit, however, because all models greatly overestimate the quiet TEC, the baseline for this parameter is unreliable and the results can be disregarded.

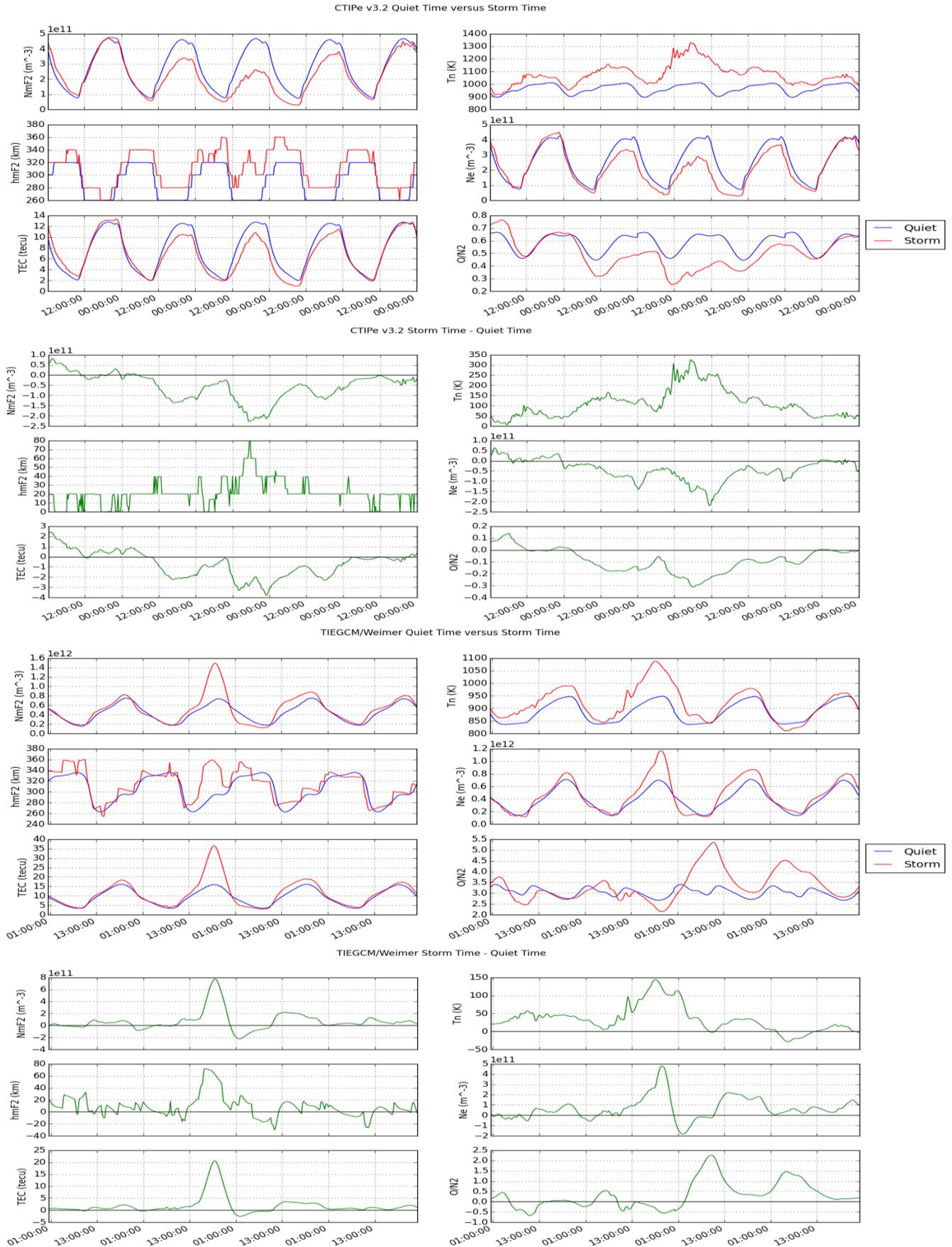

**Figure 11.** CTIPe v3.2 and TIEGCM+Weimer quiet time results (blue) compared to storm results (red). Quiet time data subtracted from storm data is shown in green. The most obvious difference between the two models is that while TIEGCM values increase during the storm, CTIPe values decrease.

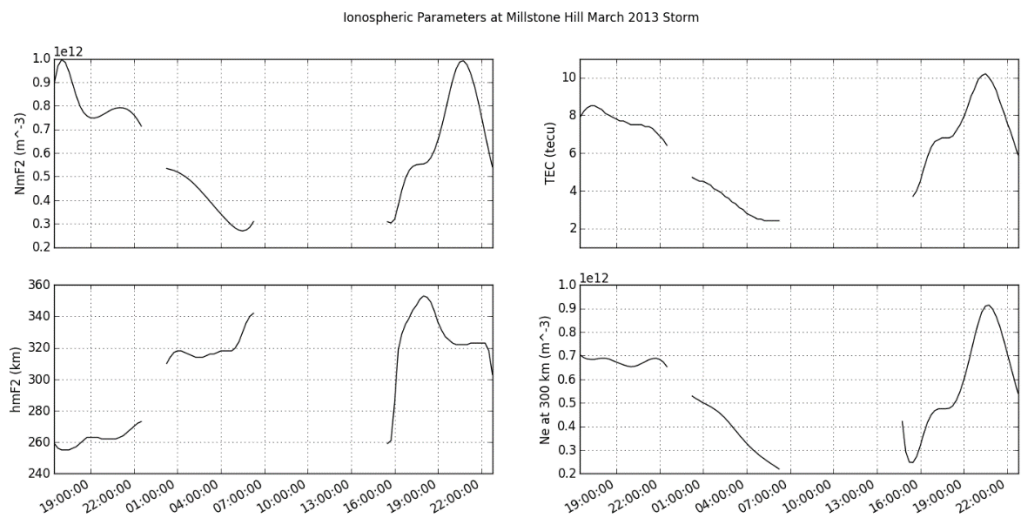

**Figure 12.** Millstone Hill data from the geomagnetic storm on March 17, 2013. Storm begins at 6 UT.

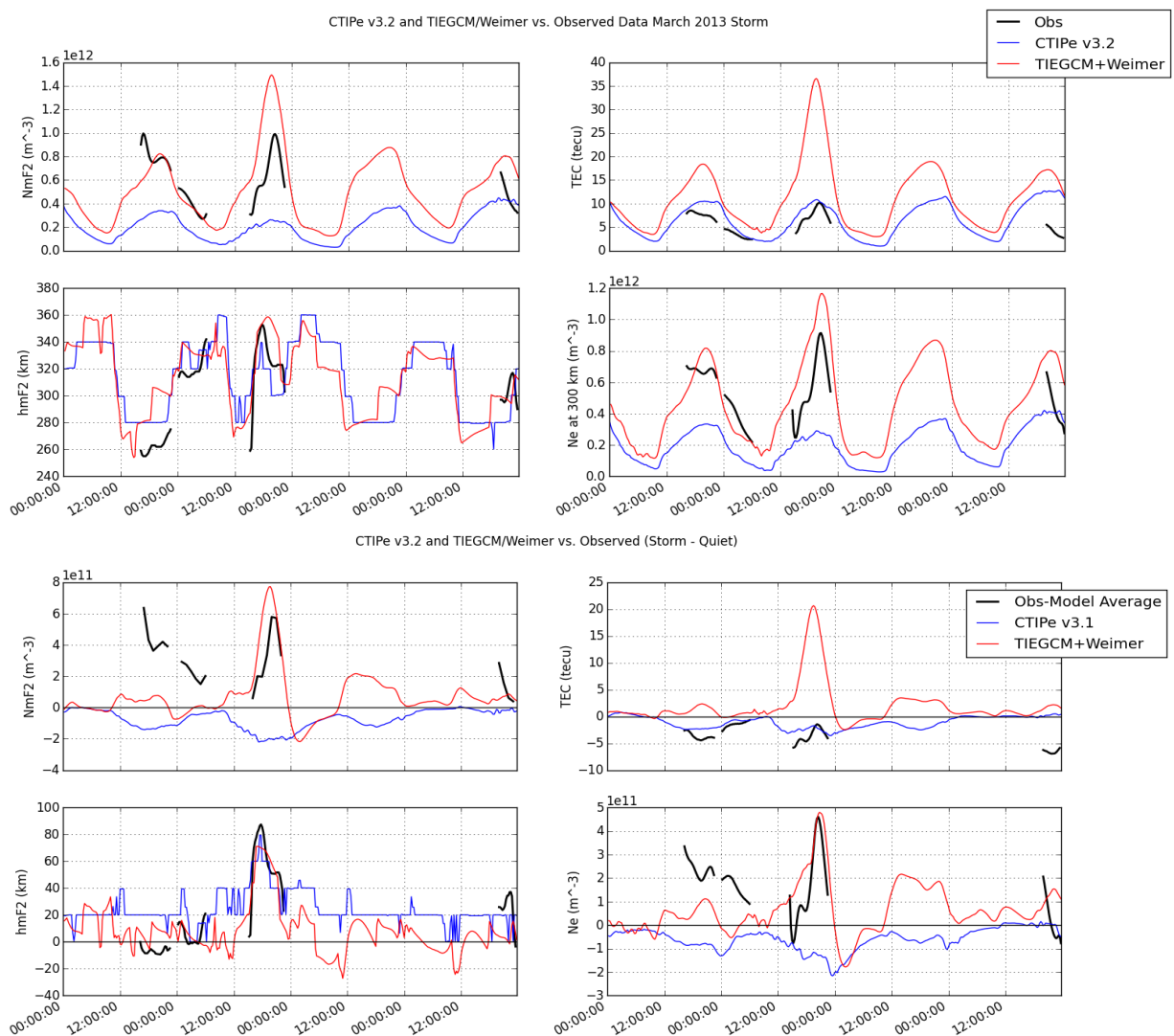

**Figure 13.** Model storm data compared to storm data. The first set of plots shows data comparisons between CTIPe (blue), TIEGCM (red), and observed (black). The second set of plots shows the comparisons between the differences of the quiet time and storm time data, using the average of all the models as the quiet time baseline for the observed data.

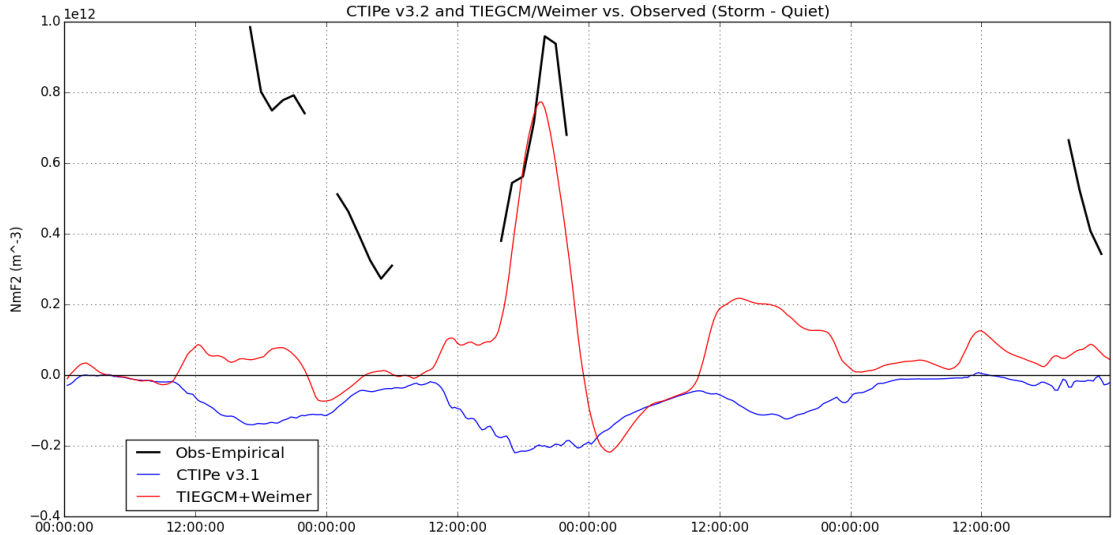

**Figure 14.** Similar to Figure 13, except using the empirical results as the quiet time baseline for the observed data. Because only NmF2 is available for the empirical model, this is the only parameter compared.

## 6 Conclusions and Future Research

We have investigated the impact of solar activity on the ionosphere during geomagnetic quiet time and how this information can be used to analyze the effects of geomagnetic storms on ionospheric parameters relating to the electron density. Our study was focused on the Millstone Hill Observatory (42.6 N, 71.5 W) during the 2013 spring equinox. Before the bulk of the research was performed, two important discoveries were made: high values of F10.7 (>180) can be found even during geomagnetic quiet time, and the model CTIPe v3.1 may require a longer warm-up period to obtain proper values.

Our analysis of the model results for different F10.7 values has shown that solar activity has the greatest effect on the model output for TEC and the least effect on output for hmF2 and O/N2. Linearity with F10.7 is true for NmF2, hmF2, TEC, and Tn at 300 km but not for Ne at 300 km or O/N2 at 300 km, especially for high flux values. Comparing the TIEGCM models to the CTIPe models found that TIEGCM results tend to be higher than CTIPe results and while F10.7 value has an increased effect on output for the CTIPe models during the local nighttime/morning, the same increased effect on output is seen during the local daytime/early evening for the TIEGCM models.

When model results were compared to observed data, CTIPe v3.2 performed the best overall out of all the models, however, TIEGCM+Weimer was slightly better for NmF2. The studied models performed very well for NmF2 and Ne at 300 km, but slightly overestimated hmF2. TEC was greatly overestimated in all models throughout the majority of the day.

The results of climatology studies such as these provide us with a quiet time baseline that can be used when analyzing the effects of geomagnetic storms on the ionosphere. An example of this was presented in Section 5. Results from CTIPe v3.2 and TIEGCM+Weimer were compared against observed data from the geomagnetic storm that occurred on March 17, 2013. The comparisons showed that, while slightly overestimating the values, TIEGCM+Weimer was best at predicting NmF2 and Ne at 300 km. Both models performed equally well when predicting hmF2. CTIPe v3.2 had the best fit for TEC, while once again TIEGCM greatly overestimated the value. When comparing the differences between quiet time values and storm values, similar trends in performance for each model can be seen.

To fully establish a reliable estimate of the impact of solar activity on the ionosphere, more research will have to be done using various latitudes, seasons, and years during the solar cycle. Additionally, as these models are updated and improved it will continue to be important to validate their results against observed data.